\documentclass[journal,comsoc]{IEEEtran}
\usepackage[T1]{fontenc}
\usepackage{algorithm,algorithmic,amsbsy,amsmath,amssymb,epsfig,bbm,mathrsfs,multirow,amsthm}
\usepackage{subcaption,float}
\usepackage[hidelinks]{hyperref}

\begin{document}

\title{Distributed Deep Learning at the Edge: A Novel Proactive and Cooperative Caching Framework for Mobile Edge Networks}

\author{\IEEEauthorblockN{Yuris Mulya Saputra, Dinh Thai Hoang, Diep N. Nguyen, Eryk Dutkiewicz, Dusit Niyato, and Dong In Kim \vspace*{-4mm}}
\thanks{Y.~M.~Saputra and D.~T.~Hoang, D.~N.~Nguyen, and E.~Dutkiewicz are with University of Technology Sydney, Australia (email: yurismulya.saputra@student.uts.edu.au, Hoang.Dinh, Diep.Nguyen, and Eryk.Dutkiewicz@uts.edu.au).}
\thanks{ D.~Niyato is with Nanyang Technological University, Singapore (email: dniyato@ntu.edu.sg).}
\thanks{D.~I.~Kim is with Sungkyunkwan University, South Korea (e-mail: dikim@skku.ac.kr).}}

\maketitle

\begin{abstract}
This letter proposes two novel proactive cooperative caching approaches using deep learning (DL) to predict users' content demand in a mobile edge caching network. In the first approach, a (central) content server takes responsibilities to collect information from all mobile edge nodes (MENs) in the network and then performs our proposed deep learning (DL) algorithm to predict the content demand for the whole network. However, such a centralized approach may disclose the private information because MENs have to share their local users' data with the content server. Thus, in the second approach, we propose a novel distributed deep learning (DDL) based framework. The DDL allows MENs in the network to collaborate and exchange information to reduce the error of content demand prediction without revealing the private information of mobile users. Through simulation results, we show that our proposed approaches can enhance the accuracy by reducing the root mean squared error (RMSE) up to $33.7$\% and reduce the service delay by $36.1$\% compared with other machine learning algorithms.
\end{abstract}


\section{Introduction}

\IEEEPARstart{M}{obile} edge caching (MEC) has been emerging as one of the most effective solutions to deal with the ever-increasing traffic demand for new services (e.g., video streaming, IoT, and virtual reality applications) in mobile networks. The key idea of an MEC network is to distribute popular contents closer to the mobile users via mobile edge nodes (MENs)~\cite{Mao:2017} to reduce the service delay for the mobile users. As a result, the deployment of MEC network helps to improve the users' experiences (e.g., trustworthy wireless connections, fast data transfer, and low energy consumption) and thus maximize the revenues for the MEC service providers~\cite{Hoang:2018}.

To efficiently cache the popular contents in the MEC network, proactive caching is one of the most effective methods to predict the mobile users' demands (i.e., content requests). In particular, the proactive caching can provide optimal caching decisions to increase the cache hit rate and reduce the operational as well as service costs on the backhaul link for the MEC service providers~\cite{Hoang:2018}. In~\cite{Zeydan:2016}, a learning based proactive caching using singular value decomposition (SVD) to cache data at the base stations was investigated. In this work, the data is first collected from the base stations and then trained in a big data platform. Nevertheless, the SVD technique sets all empty entries to be zero, leading to a poor prediction accuracy, especially when a dataset is extremely sparse. Furthermore, the SVD observes approximated ranks of elements and thus may produce negative numbers which provide no information about real users' demands. To address this problem, the authors in~\cite{Ahn:2018} adopted the non-negative matrix factorization (NMF) to predict the demand probability along with an implicit feedback of the users' social context. As such, the NMF technique applies the additive parts-based representation with non-negative elements to enhance the interpretability of the elements when the dataset is sparse. However, the NMF is a linear model which considers only two-factor correlation (i.e., the user-content relationship) without learning multi-level correlation. Given that, deep learning seems a suitable solution that relies on deep neural networks (DNN) to learn multiple levels of processing layers. Each layer of the DNN provides nonlinear transformations of the complex hidden features to obtain correlations between the mobile users and the content demands hierarchically (i.e., a layer learns and aggregates a set of features according to the previous layer's results)~\cite{Zhang4:2018}. 

In this letter, we introduce two novel proactive cooperative caching approaches using DL algorithms to improve the accuracy of content demand prediction for the MEC network. In the first approach, we develop a model utilizing the content server (CS) as a centralized node to collect information from all the MENs. We then use the DL to predict the demands for the whole network. However, such an approach may raise the concerns on information privacy and communication overhead. To address these problems, we propose the novel approach using DDL-based framework. In this framework, the CS only needs to collect the trained models from MENs and update the global model accordingly~\cite{Dean:2012}. After that, the global model will be sent back to the MENs for further updates. Through simulation results, we demonstrate that both proposed approaches can improve the accuracy of prediction up to $33.7$\% and reduce the service delay by $36.1$\% compared with other well-known proactive caching algorithms at MENs, i.e., SVD and NMF.

\section{System Model}
\label{Sec.System}

\begin{figure}[!t]
	\centering
	\includegraphics[scale=0.32]{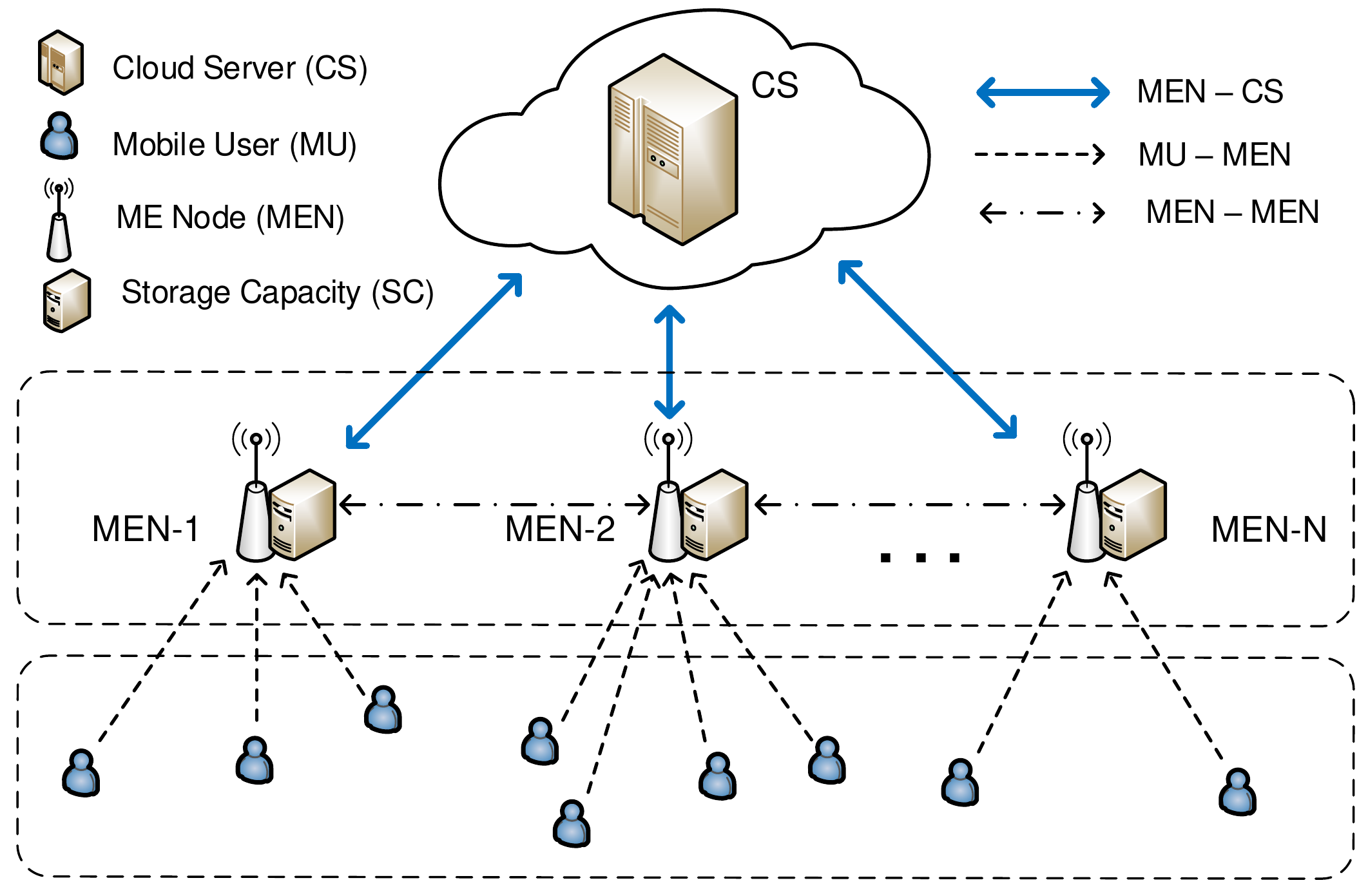}
	\caption{Network architecture.}
	\label{fig:Edge_Architecture}
\end{figure}

\subsection{Network Architecture}

The proposed network architecture is illustrated in Fig.~\ref{fig:Edge_Architecture}. Mobile users are connected to MENs within their service area. All MENs are also connected to the CS through the backhaul links by using either wireless (i.e., cellular networks) or wired connections. Each MEN is equipped with a finite storage capacity to cache popular contents locally according to the decision of proactive cooperative caching framework. When a user sends a content request to an MEN, the content will be sent to the user instantly if the content is stored locally at the MEN. Otherwise, the MEN downloads the content from the CS or from one of its directly connected MENs, and sends it to the requesting user. We denote $\mathcal{N} = \{1,\ldots,n,\ldots,N\}$ as the set of MENs, $\mathcal{U} = \{\mathcal{U}_1,\ldots,\mathcal{U}_n,\ldots,\mathcal{U}_N\}$ as the set of mobile users in the MEC network, and $U$ as the total number of mobile users in the network. In this way, $\mathcal{U}_n = \{1,\ldots,u_n,\ldots,U_n\}$ represents the set of mobile users at MEN-$n$'s coverage area. Each MEN-$n$ has the storage capacity denoted by $S_n$. Note that mobile users can move and download their requested contents from any MEN in the network. Thus, the set of users $\mathcal{U}_n$ captures all users who visit and download contents at MEN-$n$. In addition, the set $\mathcal{U}_n$ also captures the case when a user, e.g., user $u_n$, downloads a content at MEN-$n$ via another MEN. Then, we denote $\mathcal{I} = \{1,\ldots,i,\ldots,I\}$ as the set of contents.

\subsection{Proactive Cooperative Caching Mechanism}

To cache popular contents, each MEN, e.g., MEN-$n$, collects from mobile users in its serving area and sets up a dataset, i.e.,  $\mathbf{X}_n$, containing popularity factors $f^i_{u_n}$ of user $u_n$ over the content $i$. For the first approach (i.e., using DL in the CS), the CS collects $\mathbf{X}_n, \forall n \in \mathcal{N}$ from the MENs cooperatively and then concatenates them into dataset $\mathbf{X}_{cs}$ vertically with popularity factor $f^i_u$ of the user $u$ (where $u \in \mathcal{U}$) over the content $i$. In this way, MENs can share the model information to improve the prediction accuracy for the whole network. We use $\mathbf{X}_{\emph{\mbox{cs}}}$ to predict the content demands and then generate dataset $\mathbf{\hat Y}_{\emph{\mbox{cs}}}$ containing predicted popularity factors ${\hat f}^i_u$ at the CS. This $\mathbf{\hat Y}_{\emph{\mbox{cs}}}$ is then sent back to the MENs for content placement decision. Specifically, each MEN-$n$ obtains ${\hat f}^i_n = \underset{u \in \mathcal{U}}{\sum} {\hat f}^i_{u}$ as the predicted popularity factor aggregation of content $i$. 


In the second approach with the DDL, each MEN-$n$ can predict the demands locally using $\mathbf{X}_n$. Then, the CS only needs to collect the trained models from MENs and update the global model cooperatively (explained in Section~\ref{Sec.PF2}) and create $\mathbf{\hat Y}_{\emph{\mbox{n}}}$ which contains predicted popularity factors ${\hat f}^i_{u_n}$. To perform the content placement decision, each MEN-$n$ aggregates the predicted popularity factors of content $i$ as ${\hat f}^i_n = \underset{u_n \in \mathcal{U}_n}{\sum} {\hat f}^i_{u_n}$. Based on ${\hat f}^i_n$ of the first and second approaches, we can obtain specific largest numbers of ${\hat f}^i_n$ at MEN-$n$ in descending order. In particular, we select the contents with top-$R$ of ${\hat f}^i_n$ which are likely to be cached at MEN-$n$.

\section{DL-Based Proactive Cooperative Caching} 
\label{Sec.PF}

In this approach, the CS needs to learn $\mathbf{X}_{\emph{\mbox{cs}}}$ through the DNN by partitioning $\mathbf{X}_{\emph{\mbox{cs}}}$ into smaller subsets (referred to as mini-batch size $\beta$). 
For DNN, each layer $\ell$ produces an output matrix containing global weight matrix $\mathbf{W}_\ell$ to control how strong the influence of a layer's each neuron to the other, and global bias vector $\mathbf{v}_\ell$ to fit the dataset as follows:
\begin{equation}
\label{eqn1}
\begin{aligned}
\mathbf{Y}_{\emph{\mbox{cs}}}^\ell = \alpha_{\emph{\mbox{cs}}} \big(\mathbf{W}_\ell\mathbf{X}_{\emph{\mbox{cs}}}^\ell + \mathbf{v}_\ell\big),
\end{aligned}
\end{equation}
where $\mathbf{X}_{\emph{\mbox{cs}}}^\ell$ is the input matrix (i.e., training dataset) of layer $\ell$ in the CS (with $\mathbf{X}_{\emph{\mbox{cs}}}^1 = \mathbf{X}_{\emph{\mbox{cs}}}$) and $\alpha_{\emph{\mbox{cs}}}$ is the \emph{rectified linear unit (ReLU)} activation function to transform the input of the layer into a nonlinear form for learning more complex feature interaction patterns. In this case, $\alpha_{\emph{\mbox{cs}}}$ returns $\mathbf{X}_{\emph{\mbox{cs}}}^\ell$ if it receives any positive input, and zero otherwise.
As the DNN contains several layers including the hidden layers, we can express $\mathbf{X}_{\emph{\mbox{cs}}}^{\ell+1} = \mathbf{Y}_{\emph{\mbox{cs}}}^\ell$. To prevent the overfitting problem and the generalization error \cite{Srivastava:2014}, we augment a dropout layer $\ell_{\emph{\mbox{drop}}}$ just after the last hidden layer. This additional layer randomly drops the input $\mathbf{X}_{\emph{\mbox{cs}}}^{\ell_{\emph{\mbox{drop}}}}$ by a fraction rate $r$, and thus the rest of the input elements are scaled by $\frac{1}{1-r}$. Then, the output layer $L$ will generate $\mathbf{Y}_{\emph{\mbox{cs}}}^L$ which is used to find the prediction loss for each mini-batch iteration $\tau$. In particular, if we consider $\mathbf{\omega} = (\mathbf{W}, \mathbf{v})$, where $\mathbf{W} = [\mathbf{W}_1,\ldots,\mathbf{W}\ell,\ldots,\mathbf{W}_L]$ and $\mathbf{v} = [\mathbf{v}_1,\ldots,\mathbf{v}_\ell,\ldots,\mathbf{v}_L]$, as the global model for all DNN layers, the prediction loss $p(\mathbf{\omega_\tau})$ for one $\tau$ in the CS is expressed by the mean-squared error (MSE) $p(\mathbf{\omega_\tau}) = \frac{1}{\beta}{\underset{u=1}{\overset{\beta}{\sum}}} p_u(\mathbf{\omega_\tau})$, where $p_u(\mathbf{\omega_\tau}) = (y_{\emph{\mbox{cs}}}^u - x_{\emph{\mbox{cs}}}^u)^2$. Here, $x_{\emph{\mbox{cs}}}^u$ and $y_{\emph{\mbox{cs}}}^u$ are the elements of matrices $\mathbf{X}_{\emph{\mbox{cs}}}^1$ and $\mathbf{Y}_{\emph{\mbox{cs}}}^L$, respectively. Then, we can compute the global gradient of using DL by $G_{\tau} = \nabla \mathbf{\omega}_\tau = \frac{\partial p(\mathbf{\omega_\tau})}{\partial \mathbf{\omega}_\tau}$.

After $G_\tau$ is obtained, the CS updates the global model $\mathbf{\omega}_{\tau}$ with the aim to minimize the prediction loss function, i.e.,  $\underset{\mathbf{\omega}}{\text{\bf min }}p(\mathbf{\omega})$. As such, we adopt the adaptive learning rate optimizer \emph{Adam} to provide fast convergence and profound robustness to the model~\cite{Kingma:2015}. Consider $\eta_\tau$ and $\delta_\tau$ to be the exponential moving average (to estimate the mean) of the $G_\tau$  and the squared $G_\tau$ to predict the variance at $\tau$, respectively. Then, the update rules of $\eta_{\tau+1}$ and $\delta_{\tau+1}$ can be expressed by:
\begin{equation}
\label{eqn6}
\eta_{\tau+1} = \gamma_\eta^\tau \eta_{\tau} + (1 - \gamma_\eta^\tau)G_\tau, \mbox{and } \delta_{\tau+1} = \gamma_\delta^\tau \delta_{\tau} + (1 - \gamma_\delta^\tau)G_\tau^2,
\end{equation}
where $\gamma_\eta^\tau$ and $\gamma_\delta^\tau \in [0,1)$ represent the exponential decay steps of $\eta_\tau$ and  $\delta_\tau$ at $\tau$, respectively. To update the global model, we also consider the learning step $\lambda$ to decide how fast the global model will be updated at each $\tau$. In particular, the update rule for $\lambda$ follows this expression:
\begin{equation}
\label{eqn7}
\begin{aligned}
\lambda_{\tau+1} = \lambda\frac{\sqrt{1 - \gamma_\delta^{\tau+1}}}{1 - \gamma_\eta^{\tau+1}}.
\end{aligned}
\end{equation}
Then, the global model $\mathbf{\omega}_{\tau+1}$ for the next $\tau+1$ is updated by:
\begin{equation}
\label{eqn8}
\begin{aligned}
\mathbf{\omega}_{\tau+1} = \mathbf{\omega}_{\tau} - \lambda_{\tau+1}\frac{\eta_{\tau+1}}{\sqrt{\delta_{\tau+1}} + \epsilon},
\end{aligned}
\end{equation}
where $\epsilon$ indicates a constant to avoid zero division when the $\sqrt{\delta_{\tau+1}}$ is almost zero. For this approach, $\mathbf{\omega}_{\tau+1}$ is used to learn the dataset for the next $\tau+1$ in the CS. The same process is repeated until each sample $u$ of $\mathbf{X}_{\emph{\mbox{cs}}}$ has been observed referred to as epoch time $t$. Then, the process terminates when the prediction loss converges or the certain number of epoch time $T$ is reached. In this case, we can obtain the final global model $\mathbf{\omega}^*$ to predict $\mathbf{\hat Y}_{\emph{\mbox{cs}}}$ of training dataset $\mathbf{X}_{\emph{\mbox{cs}}}$ and new dataset $\mathbf{\hat X}_{\emph{\mbox{cs}}}$ using Eq.~(\ref{eqn1}). The algorithm for proactive cooperative caching using DL is shown in Fig.~\ref{DDL-PC} in which the process inside the dotted block (A) is executed at the CS.

\section{DDL-Based Proactive Cooperative Caching} 
\label{Sec.PF2}


In this approach, each MEN distributedly implements the DL technique to learn from its dataset $\mathbf{X}_{n}$ locally. The $\mathbf{X}_{n}$ is then divided into smaller subsets with mini-batch size $\frac{\beta}{N}$. For DNN, each MEN-$n$ generates the output matrix 
\begin{equation}
\label{eqn1b}
\begin{aligned}
\mathbf{Y}_n^\ell = \alpha_n \big(\mathbf{W}_\ell\mathbf{X}_n^\ell + \mathbf{v}_\ell\big),
\end{aligned}
\end{equation}
where $\mathbf{X}_n^\ell$ is the input matrix of layer $\ell$ at MEN-$n$ (with $\mathbf{X}_{\emph{\mbox{n}}}^1 = \mathbf{X}_{\emph{\mbox{n}}}$) and $\alpha_n$ is the \emph{ReLU} activation function at MEN-$n$. We also drop the input $\mathbf{X}_n^{\ell_{\emph{\mbox{drop}}}}$ in the dropout layer by a fraction rate $r$. In the output layer, we can generate $\mathbf{Y}_n^L$ and find the prediction loss for each $\tau$ by
$p_n(\mathbf{\omega_\tau}) = \frac{N}{\beta}{\underset{u=1}{\overset{\frac{N}{\beta}}{\sum}}} p_n^u(\mathbf{\omega_\tau})$, where $p_n^u(\mathbf{\omega_\tau}) = (y_n^u - x_n^u)^2$. Here, $x_n^u$ and $y_n^u$ are the element of matrices $\mathbf{X}_n^1$ and $\mathbf{Y}_n^L$ at MEN-$n$, respectively.  Next, we can compute the local gradient by $g_n^\tau = \nabla \mathbf{\omega}_\tau = \frac{\partial p_n(\mathbf{\omega_\tau})}{\partial \mathbf{\omega}_\tau}$. When $g_n^\tau$ computation is completed for each $\tau$, each MEN will send this local gradient to the CS for global gradient aggregation $G_\tau$. Specifically, the CS acts as a parameter server to aggregate the gradients of the models from all connected MENs and then update the global model $\mathbf{\omega}_\tau$ by using Eq.~(\ref{eqn8}) before sending back to the MENs. Doing so allows all MENs to collaborate by sharing local model information to each other to further improve the prediction accuracy through the CS. To guarantee that the gradient staleness is 0, the gradient averaging process is enabled right after $N$ local gradients, i.e., $g_n^\tau$ are received by the CS synchronously. Here, the gradient staleness happens when local gradients are computed using an obsolete/non-latest global model. Then, the global gradient $G_\tau$ of the DDL is $G_\tau = \frac{1}{N}{\underset{n=1}{\overset{N}{\sum}}} g_n^\tau$. 

\begin{figure}[!t]
	\centering
	\includegraphics[scale=0.5]{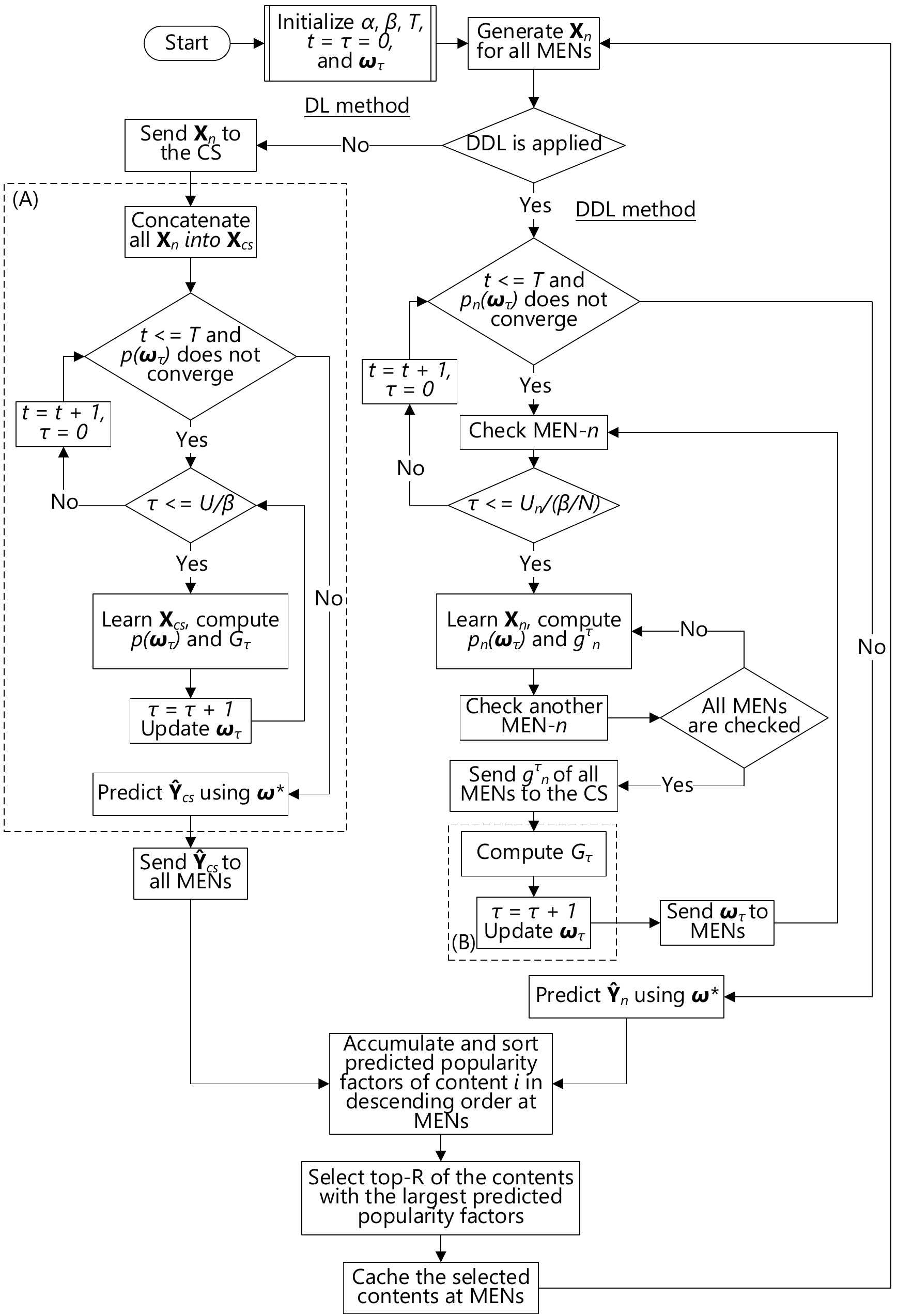}
	\caption{Flowchart of DL and DDL approaches.}
	\label{DDL-PC}
\end{figure}

To minimize the prediction loss function, i.e.,  $\underset{\mathbf{\omega}}{\text{\bf min }}p_n(\mathbf{\omega})$, at each MEN-$n$, we also adopt the Adam optimizer and update the global model $\mathbf{\omega}_{\tau+1}$ as expressed in Eqs.~(\ref{eqn6})-(\ref{eqn8}). This $\mathbf{\omega}_{\tau+1}$ is then sent back to the MENs for the next local learning process. The aforementioned process continues until the prediction loss converges or $T$ is reached. We then can predict $\mathbf{\hat Y}_{\emph{\mbox{n}}}$ of training dataset $\mathbf{X}_n$ and new dataset $\mathbf{\hat X}_n$ at each MEN using $\mathbf{\omega}^*$ through Eq.~(\ref{eqn1b}). The algorithm for proactive cooperative caching using DDL is summarized in Fig.~\ref{DDL-PC}. The process inside the dotted block (B) is implemented at the CS.

\section{Performance Evaluation} 


\begin{figure*}[htbp]
	\centering
	\begin{subfigure}[b]{0.33\textwidth}
		\centering
		\includegraphics[scale=0.4]{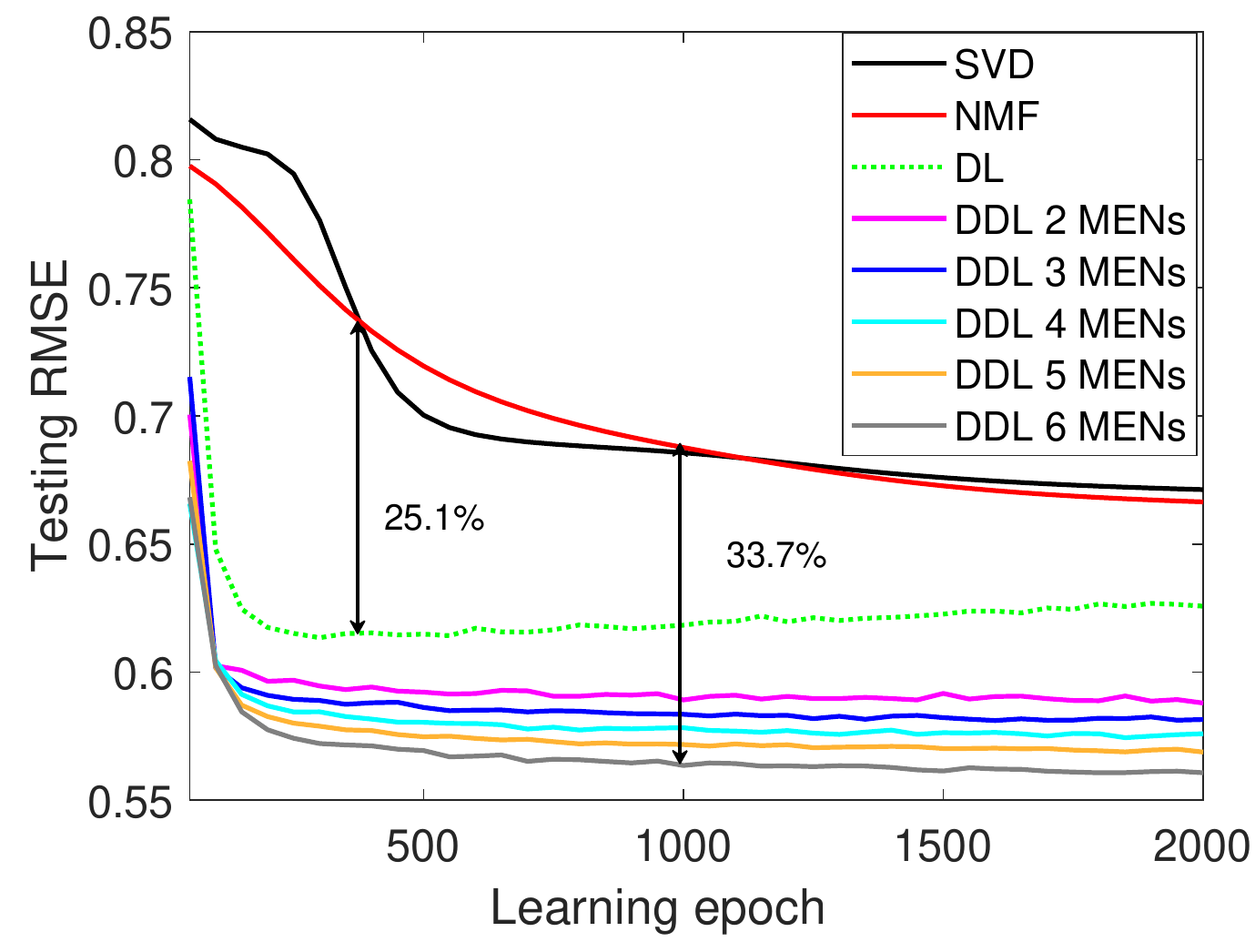}
		\caption{RMSE vs prediction methods}
	\end{subfigure}%
	~ 
	\begin{subfigure}[b]{0.33\textwidth}
		\centering
		\includegraphics[scale=0.4]{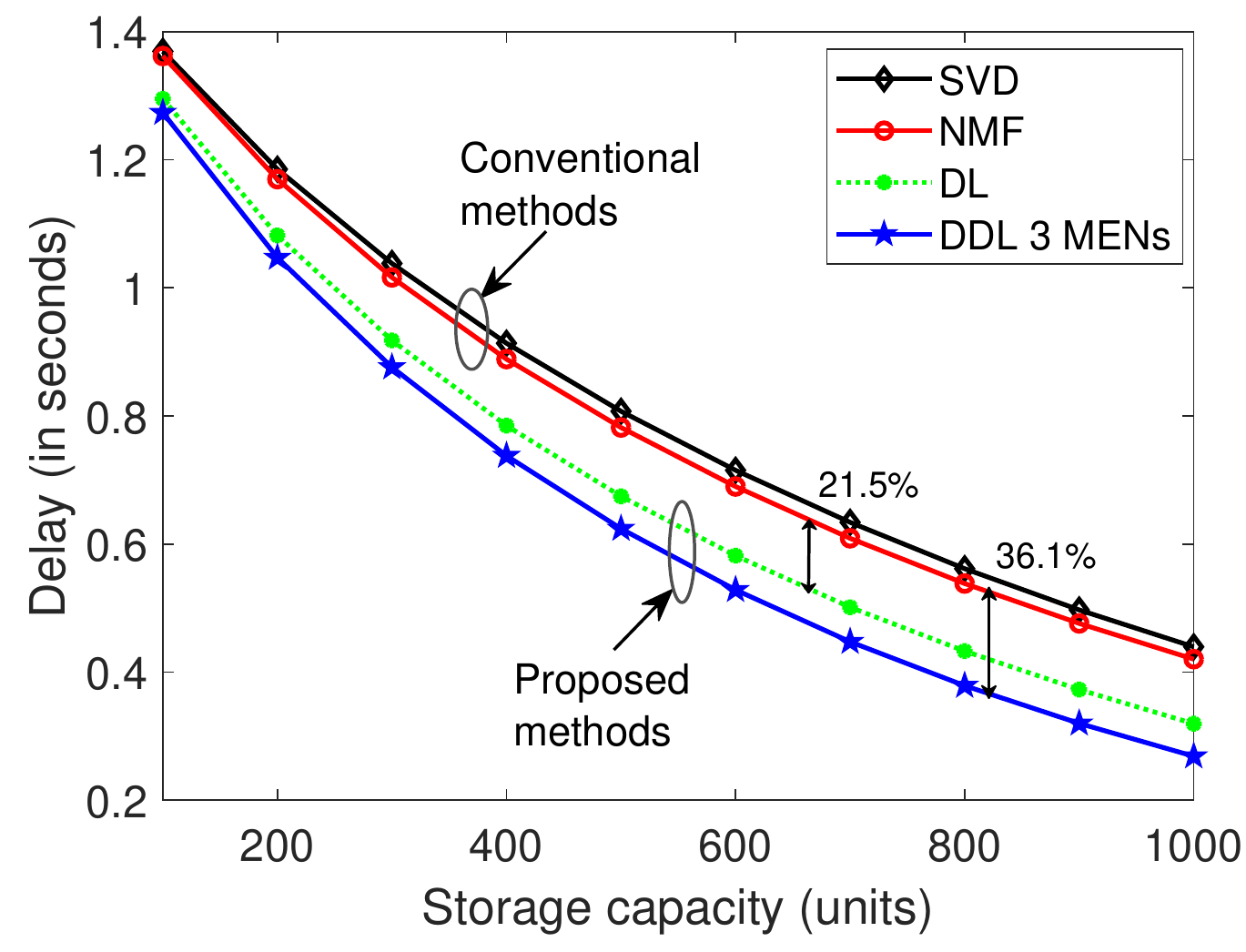}
		\caption{Average delay vs storage capacity}
	\end{subfigure}%
	~ 
	\begin{subfigure}[b]{0.33\textwidth}
		\centering
		\includegraphics[scale=0.4]{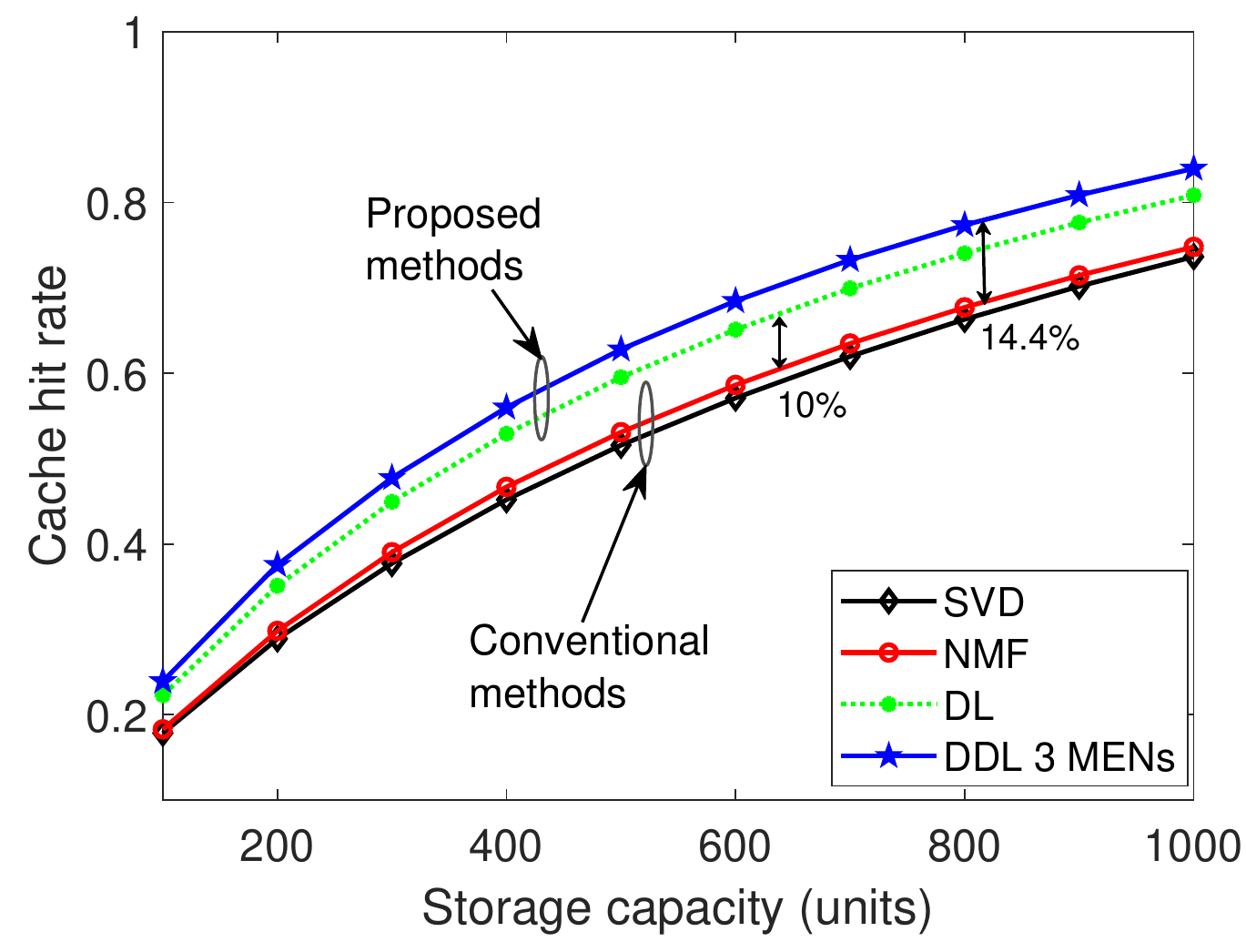}
		\caption{Cache hit rate vs storage capacity}
	\end{subfigure}
	\caption{The performance comparison of SVD, NMF, DL, and DDL prediction methods.} 
	\label{fig:performance}
\end{figure*}

\begin{figure}[!t]
	\centering
	\includegraphics[scale=0.4]{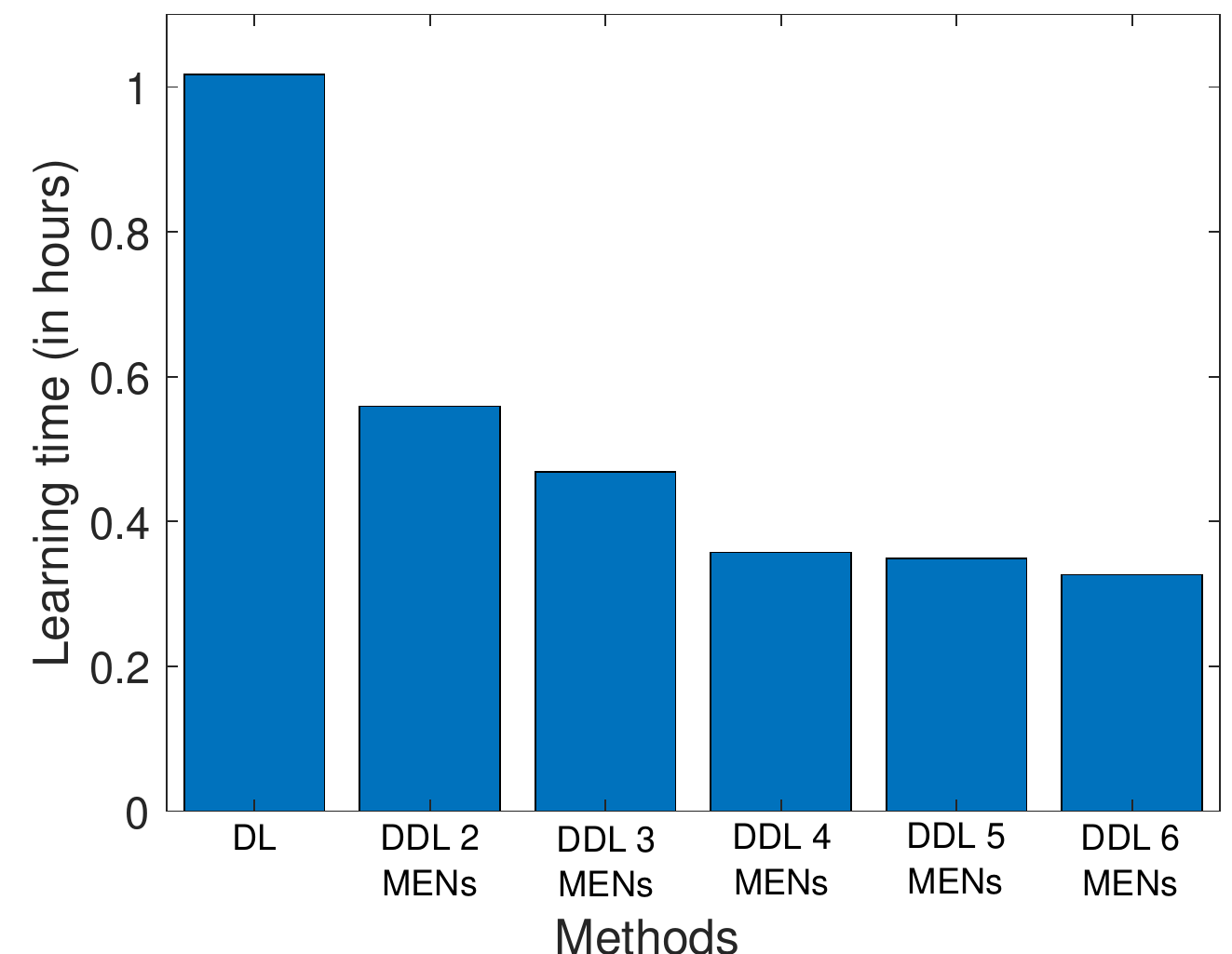}
	\caption{Learning 2000 epoch time for various methods.}
	\label{fig:Learning_time}
\end{figure}

\subsection{Experimental Setup} 

We evaluate the performance of the content prediction on the MEC network with one CS and six MENs using \emph{TensorFlow CPU} library in Linux platform of an Intel Xeon Gold 6150 2.7GHz 18 cores with 180GB RAM. We compare our proposed frameworks with two well-known machine learning methods including SVD~\cite{Zeydan:2016} and NMF~\cite{Mao:2006}. We use Movielens 1M-dataset with more than 1M ratings from 6040 users with 3952 movies. Then, we split the dataset into 80\% training dataset and 20\% testing dataset. From the training dataset, we divide the number of samples equally with respect to the number of MENs when DDL is implemented. Each MEN runs the testing dataset for the popularity factor prediction to compute the performance metrics. For DNN, we use two hidden layers with $64$ neurons per layer and one dropout layer with a fraction rate $0.8$. We also apply the adaptive learning rate optimizer Adam with initial step size $0.001$ and $2000$ epoch time during the learning process. Furthermore, for the content placement algorithm, we consider the same size for each content at 200MB. The bandwidth between an MEN and the CS is set at 60Mbps.

\subsection{Simulation Results}
Fig.~\ref{fig:performance} shows the comparison between conventional baseline and proposed methods. We first evaluate the prediction accuracy, i.e., RMSE, as the learning epoch increases in Fig.~\ref{fig:performance}(a). In particular, the RMSE obtained by the DDL is 33.7\% lower than those of SVD and NMF. The reason is that DDL can deeply learn the meaningful features from the subset of the whole dataset independently at different MENs, and thus the sensitivity to learn new testing dataset becomes better when the local models obtained by the MENs are aggregated together. In other words, the average prediction of all MENs will produce less variance and lower error regarding the number of MENs \cite{Guo:1999}. In contrast, SVD and NMF only generate linear assumptions of two factors based on the low-rank approximation~\cite{Mao:2006} without deeply learning the representations, and thus the RMSE cannot be minimized properly. For the DL, although the RMSE is higher than that of the DDL, the DL can improve the RMSE by as much as 25.1\% compared with those of the SVD and NMF.

We then observe the average delay to download contents from the CS and cache hit rate when the storage capacity increases in Figs.~\ref{fig:performance}(b) and~\ref{fig:performance}(c), respectively. Align with the trend of the RMSE, the DL and DDL approaches can reduce average delay up to 21.5\% and 36.1\% and increase the cache hit rate by 10\% and 14.4\%, respectively, compared with those of SVD and NMF. The reason is that the proposed approaches can optimize the use of hyperparameter settings to improve the accuracy of content demand prediction. Examples of the hyperparameters settings include the number of hidden layers and neurons, the regularization methods, the activation functions, and the size of mini-batch. We also observe in Fig.~\ref{fig:Learning_time} that the DDL can learn the dataset faster than the DL as the number of the MENs increases. This interesting trend can provide useful information for MEC service providers to tradeoff between the learning time of the users' demands and the implementation costs in the MEC network.

\section{Summary} 

In this letter, we have presented two novel proactive cooperative caching approaches leveraging deep learning (DL) and distributed deep learning (DDL) algorithms for the MEC network. In the first approach, the CS collects the information from all MENs and uses the DL technique to predict the users' demands for the network. Then, to further minimize the communication overhead and address the privacy concern, we proposed the DDL-based scheme in which the DL can be executed at the edge. This scheme allows MENs to only exchange the gradient information, not the complete information of the users, and perform the DL to predict users demand without revealing the private information of the mobile users.


\end{document}